\documentclass[prl,a4paper,twocolumn,10pt,headsepline,superscriptaddress]{revtex4-2}
\pdfoutput=1

\usepackage[nouppercase]{scrlayer-scrpage}
\usepackage{graphicx}
\usepackage[left=1.57cm, right=1.57cm, top=1.8cm, bottom=2.5cm]{geometry}
\usepackage{textcomp}
\usepackage{microtype}
\usepackage{stickstootext}
\usepackage{placeins}
\usepackage{amsthm}
\usepackage[stickstoo,vvarbb]{newtxmath}
\usepackage[colorlinks=true, linkcolor=black, anchorcolor=black, citecolor=black, filecolor=black, urlcolor=black]{hyperref}

\clearscrheadfoot

\begin{document}
\title {
\large
Kinematically complete experimental study of Compton scattering at helium atoms near the ionization threshold
}
\author{Max Kircher} \email{kircher@atom.uni-frankfurt.de} \address{Institut f\"ur Kernphysik, J. W. Goethe Universit\"at, Max-von-Laue-Str. 1, D-60438 Frankfurt, Germany}
\author{Florian Trinter} \address{FS-PETRA-S, Deutsches Elektronen-Synchrotron DESY, Notkestr. 85, D-22607 Hamburg, Germany} \address{Molecular Physics, Fritz-Haber-Institut der Max-Planck-Gesellschaft, Faradayweg 4-6, D-14195 Berlin, Germany}
\author{Sven Grundmann} \address{Institut f\"ur Kernphysik, J. W. Goethe Universit\"at, Max-von-Laue-Str. 1, D-60438 Frankfurt, Germany}
\author{Isabel Vela-Perez} \address{Institut f\"ur Kernphysik, J. W. Goethe Universit\"at, Max-von-Laue-Str. 1, D-60438 Frankfurt, Germany}
\author{Simon Brennecke} \address{Institut f\"ur Theoretische Physik, Leibniz Universit\"at Hannover, Appelstr. 2, D-30167 Hannover, Germany}
\author{Nicolas Eicke} \address{Institut f\"ur Theoretische Physik, Leibniz Universit\"at Hannover, Appelstr. 2, D-30167 Hannover, Germany}
\author{Jonas Rist} \address{Institut f\"ur Kernphysik, J. W. Goethe Universit\"at, Max-von-Laue-Str. 1, D-60438 Frankfurt, Germany}
\author{Sebastian Eckart} \address{Institut f\"ur Kernphysik, J. W. Goethe Universit\"at, Max-von-Laue-Str. 1, D-60438 Frankfurt, Germany}
\author{Salim Houamer} \address{LPQSD, Department of Physics, Faculty of Science, University S\'etif-1, 19000, Setif, Algeria}
\author{Ochbadrakh Chuluunbaatar} \address{Joint Institute for Nuclear Research, Dubna, Moscow region 141980, Russia}
\address{Institute of Mathematics and Digital Technologies, Mongolian Academy of Sciences, 13330, Ulaanbaatar, Mongolia}
\address{Peoples' Friendship University of Russia (RUDN University), 117198, Moscow, Russia}
\author{Yuri~V. Popov} \address{Skobeltsyn Institute of Nuclear Physics, Lomonosov Moscow State University, Moscow 119991, Russia}\address{Joint Institute for Nuclear Research, Dubna, Moscow region 141980, Russia}
\author{Igor~P. Volobuev} \address{Skobeltsyn Institute of Nuclear Physics, Lomonosov Moscow State University, Moscow 119991, Russia}
\author{Kai Bagschik} \address{FS-PETRA-S, Deutsches Elektronen-Synchrotron DESY, Notkestr. 85, D-22607 Hamburg, Germany}
\author{M.~Novella Piancastelli} \address{Sorbonne Universit\'es, CNRS, UMR 7614, Laboratoire de Chimie Physique Mati\`ere et Rayonnement, F-75005 Paris, France} \address{Department of Physics and Astronomy, Uppsala University, P.O. Box 516, SE-751~20 Uppsala, Sweden}
\author{Manfred Lein} \address{Institut f\"ur Theoretische Physik, Leibniz Universit\"at Hannover, Appelstr. 2, D-30167 Hannover, Germany}
\author{Till Jahnke} \address{Institut f\"ur Kernphysik, J. W. Goethe Universit\"at, Max-von-Laue-Str. 1, D-60438 Frankfurt, Germany}
\author{Markus~S. Sch\"offler} \address{Institut f\"ur Kernphysik, J. W. Goethe Universit\"at, Max-von-Laue-Str. 1, D-60438 Frankfurt, Germany}
\author{Reinhard D\"orner} \email{doerner@atom.uni-frankfurt.de}
\address{Institut f\"ur Kernphysik, J. W. Goethe Universit\"at, Max-von-Laue-Str. 1, D-60438 Frankfurt, Germany}

\begin{abstract}
\noindent
Compton scattering is one of the fundamental interaction processes of light with matter. When discovered \cite{1}, it was described as a billiard-type collision of a photon `kicking' a quasi-free electron. With decreasing photon energy, the maximum possible momentum transfer becomes so small that the corresponding energy falls below the binding energy of the electron. In this regime, ionization by Compton scattering becomes an intriguing quantum phenomenon. Here, we report on a kinematically complete experiment studying Compton scattering off helium atoms in that regime. We determine the momentum correlations of the electron, the recoiling ion and the scattered photon in a coincidence experiment based on cold target recoil ion momentum spectroscopy, finding that electrons are not only emitted in the direction of the momentum transfer, but that there is a second peak of ejection to the backward direction. This finding links Compton scattering to processes such as ionization by ultrashort optical pulses \cite{2}, electron impact ionization \cite{3,4}, ion impact ionization \cite{5,6}, and neutron scattering \cite{7}, where similar momentum patterns occur.
\end{abstract}

\maketitle

\begin{figure}[t!]
	\centering
	\includegraphics{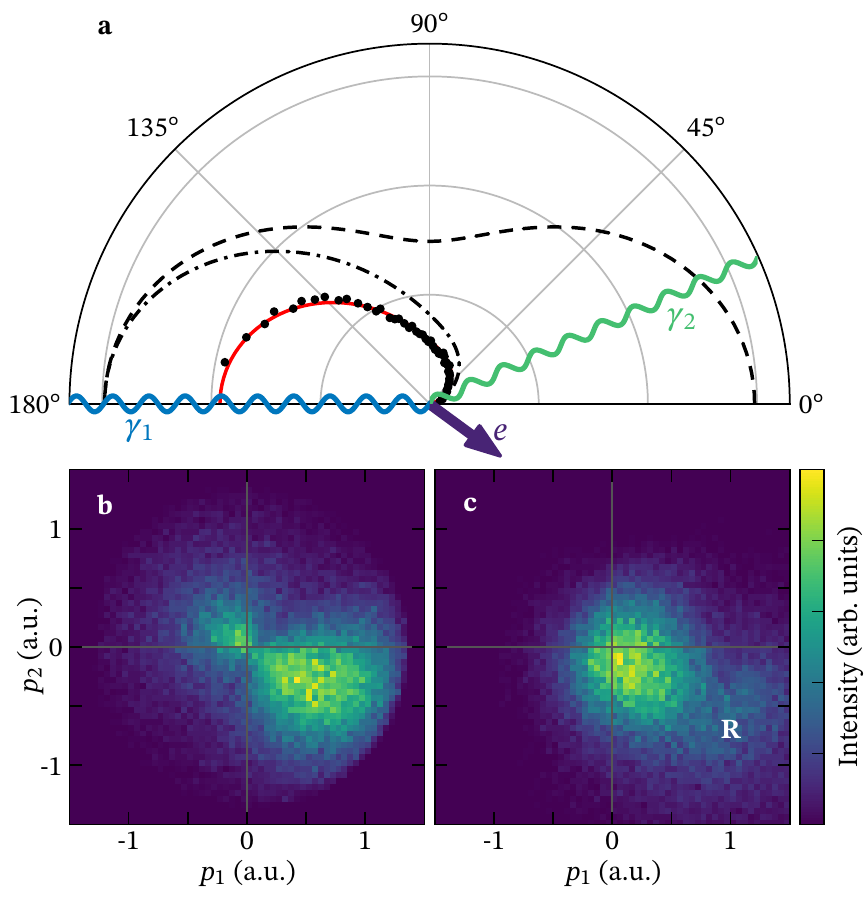}
	\caption{\small\textbf{Scheme of ionization by Compton scattering at $\boldsymbol{h\nu=2.1}$\,keV.} \textbf{a}, The wavy lines indicate the incoming and outgoing photon, and the purple arrow depicts the momentum vector of the emitted electron. The dashed line shows the Thomson cross-section, that is, the angular distribution of a photon scattering at a free electron. Black dots show the experimental photon angular distribution for ionization of He by Compton scattering, integrated over all electron emission angles and energies below 25\,eV. The photon momenta are determined using the electron and ion momenta, as well as momentum conservation. The statistical error is smaller than the dot size. The dash-dotted line shows the $A^2$ approximation for all electron energies and the solid red line shows the $A^2$ approximation for electron energies below 25\,eV. The calculations were done using Approach I (see Methods). The solid and dash-dotted lines are multiplied by a factor of 1.9. \textbf{b}, Momentum distribution of electrons emitted by Compton scattering of 2.1\,keV photons at He. The coordinate frame is the same as in \textbf{a}: the scattering plane is defined by the incoming (horizontal) and scattered photon (upper half plane); that is, $\boldsymbol p_1$ is the electron momentum component in the $\boldsymbol k_1$ direction and $\boldsymbol p_2$ is the component perpendicular to $\boldsymbol k_1$ within the scattering plane. The momentum transfer points to the forward lower half plane. The data are integrated over the out-of-plane electron momentum components. \textbf{c}, He$^+$ ion momentum distribution for the same conditions as in \textbf{b}. See main text for an explanation of the feature {\bf\textsf{R}}.}
	\label{fig1}
\end{figure}

\noindent
Doubts about energy conservation in Compton scattering at the single-event level motivated the invention, by Bothe and Geiger \cite{8}, of coincidence measurement techniques. This historic experiment settled the dispute about the validity of conservation laws in quantum physics by showing that, for each scattered photon, there is an electron ejected in coincidence. Surprisingly, however, even 95 years after this pioneering work, coincidence experiments on the Compton effect are extremely scarce and are restricted to solid-state systems \cite{9,10}. To a large extent, this lack of detailed experiments left further progress in the field of Compton scattering to theory. Due to missing experimental techniques, much of the potential of using Compton scattering as a tool in molecular physics remained untapped \cite{11}. The small cross-section of $10^{-24}$\,cm$^2$ (six orders of magnitude below typical photoabsorption cross-sections at the respective thresholds), together with the small collection solid angle of typical photon detectors, has so far prohibited coincidence experiments on free atoms and molecules. In the present work, we have solved this problem by using the highly efficient cold target recoil ion momentum spectroscopy (COLTRIMS) technique \cite{12} to detect the electron and ion momentum in coincidence. The He$^+$ ion and electrons with an energy smaller than 25\,eV are detected with $4\pi$ collection solid angle. The momentum vector of the scattered photon can be obtained using momentum conservation, thereby circumventing the need for a photon detector. This allows us to obtain a kinematically complete dataset of ionization by Compton scattering of atoms, addressing the intriguing low-energy, near-threshold regime. It has often been pointed out in the theoretical literature that such complete measurements of the process\,--\,as opposed to detection of the emitted electron or scattered photon only\,--\,are the essential key to sensitive testing of theories \cite{13} as well as allowing for a clean physics interpretation of the results \cite{14}.\\
\indent
For the case of Compton scattering at a quasi-free electron, the angular distribution of the scattered photon is given by the Thomson cross-section (Fig.~\ref{fig1}a). Binding of the electron modifies the binary scattering scenario by adding the ion as a third particle. The often invoked impulse approximation accounts for one of the effects of that binding, namely the electron's initial momentum distribution. According to this approximation, the initial electron momentum is added to the momentum balance, while the binding energy is neglected. In this model, the ion momentum is defined such that it compensates only for the electron's initial momentum. The impulse approximation works well when the binding energy is negligible compared to the energy of the electron carrying the momentum $\boldsymbol Q$ transferred by the photon. The maximum value of $Q$ is reached for photon backscattering, and is twice the photon momentum $E_1/c$, where $E_1$ is the incoming photon energy. For helium with a binding energy of 24.6\,eV, this gives a threshold of $E_1$$\approx$$\,2.5$\,keV, below which photon backscattering at an electron at rest does not provide enough energy to overcome the ionization threshold. In the present experiment, we use a photon energy of $E_1$$=$$\,2.1$\,keV, well below that threshold. Accordingly, the cross-section for ionization by Compton scattering has dropped to $\sim$20\% of
its maximum value of $\sim$10$^{-24}$\,cm$^2$ (ref. \cite{15}). As expected, we observe that the photon scattering angular distribution differs significantly from the Thomson cross-section (Fig.~\ref{fig1}a). The most striking difference is that all forward angles of photon emission are suppressed and it is almost only backscattered photons that lead to ionization. This measured cross-section shows excellent agreement with our theoretical model, which is described in detail in the Methods.

What is the mechanism facilitating ionization at these low photon energies and small momentum transfers? Our coincidence experiment can answer this question by providing the momentum vectors of all particles, that is, the incoming ($\boldsymbol k_1$) and outgoing ($\boldsymbol k_2$) photon, electron ($\boldsymbol p_e$) and ion ($\boldsymbol p_{ion}$) momentum vectors for each individual Compton ionization event. This event-by-event momentum correlation gives access to the various particles' momentum distributions in the intrinsic coordinate frame of the process, which is a plane spanned by the wavevectors of the incoming and scattered photon (Fig.~\ref{fig1}). This plane also contains the momentum transfer vector $\boldsymbol Q=\boldsymbol k_1- \boldsymbol k_2$. In Fig.~\ref{fig1}b,c, by definition, the photon is scattered to the upper half plane and the momentum transfer $\boldsymbol Q$ (that is, the `kick' by the photon) points forward and into the lower half plane. The electron momentum distribution visualized in this intrinsic coordinate frame shows two distinct islands, one in the direction of the momentum transfer and a second smaller one to the backward direction, that is, opposite to the momentum transfer direction. These two maxima are separated by a minimum. The He$^+$ ions (Fig.~\ref{fig1}c) are also emitted to the forward direction. In addition to a main island close to the origin, ions are also emitted strongly in the forward direction, towards the region indicated by {\bf{R}} (Recoil) in Fig.~\ref{fig1}c. This ion momentum distribution shows strikingly that in the below-threshold regime, the situation is very different from the quasi-free electron scattering considered in the standard high-energy Compton process. In the latter case, the ion is only a passive spectator to the photon--electron interaction and, consequently, the ion momenta are centered at the origin of the coordinate frame used in Fig.~\ref{fig1}b,c \cite{15,16,17,18}.

\begin{figure*}
	\centering
	\includegraphics{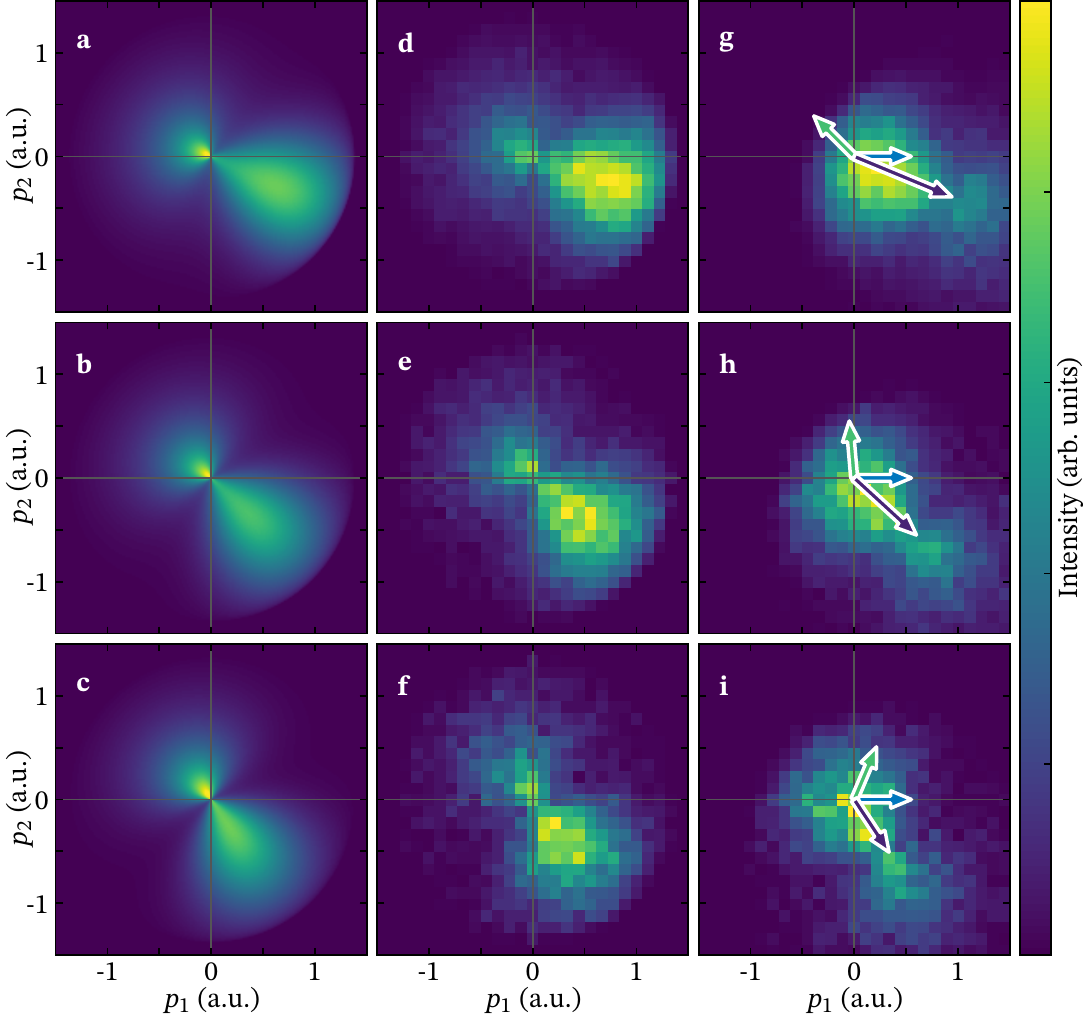}
	\caption{\small\textbf{Electron and ion momentum distributions for different momentum transfer gates.} In all panels, $p_1$ is the momentum component in $\boldsymbol k_1$ direction, $p_2$ is the component perpendicular to $\boldsymbol k_1$ within the scattering plane. \textbf{a-c}, Electron momentum distributions obtained from modeling within the $A^2$ approximation using Approach II (see Methods). \textbf{d-f}, Electron momentum distributions measured by our experiment. \textbf{g-i}, Measured momentum distributions of the ions. From top to bottom, the rows correspond to different momentum transfers $Q=1.0$, 0.8, and 0.6\,a.u., respectively. The arrows in the third column indicate the photon momentum configuration for each row. Here, the blue arrows represent the momentum of the incoming photon, the light green arrows the momentum of the scattered photon, and the dark purple arrows the momentum transfer.}
	\label{fig2}
\end{figure*}

The observed bimodal electron momentum distribution becomes even clearer when we examine a subset of the data for which the photon is scattered to a certain direction (Fig.~\ref{fig2}). This shows that the momentum distribution follows the direction of momentum transfer and the nodal plane is perpendicular to $\boldsymbol Q$. Such bimodal distributions are known from different contexts. For example, for ionization by electron impact $(e,2e)$ \cite{4} and ion impact \cite{5}, the forward lobe has been termed a binary lobe, for obvious reasons, while the backward peak is referred to as the recoil peak. This latter name alludes to the fact that, for the electron to be emitted in a direction opposite the momentum transfer, momentum conservation dictates that the ion recoils in the opposite direction. Mechanistically, this would occur if the electron was initially kicked in the forward direction but then back-reflected at its own parent ion. Such a classical picture would suggest that the ion receives the momentum originally imparted to the electron (that is, $\boldsymbol Q$) minus the final momentum, $p_e$, of the electron. This expectation is verified by our measured ion momentum distributions (Fig.~\ref{fig2}g--i). The ions also show a bimodal momentum distribution, with the main island slightly forward shifted and a minor island significantly forward shifted in the momentum transfer direction, in nice agreement with the back-reflection scheme.
\begin{figure*}
	\centering
	\includegraphics{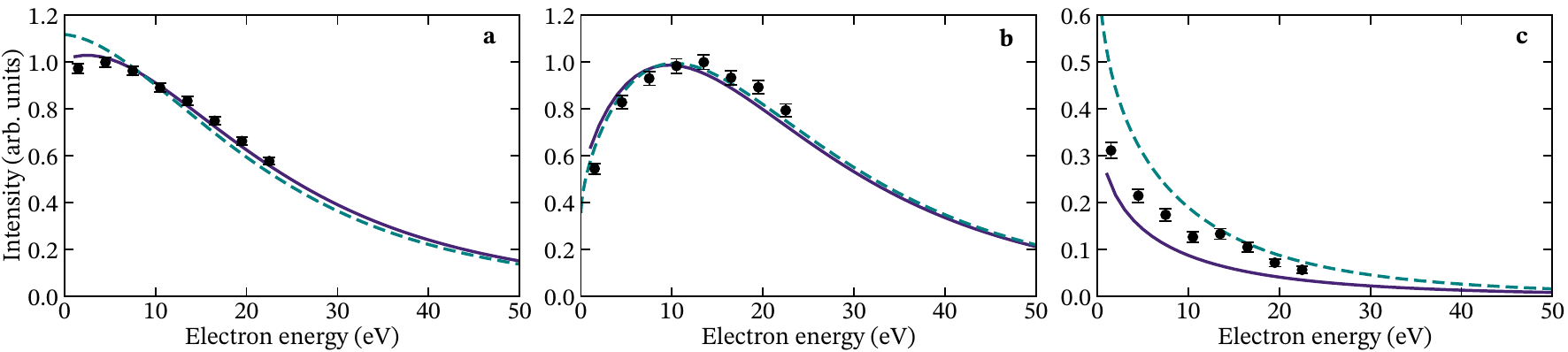}
	\caption{\small\textbf{Electron energy distribution.} The scattering angle between the incoming and outgoing photon for the outgoing photon is restricted to {$140<\theta<180$\textdegree} in all panels. \textbf a, The electron energy spectrum is shown independent of the electron emission direction. \textbf b, The electron emission angle is restricted to forward scattering ($0<\theta_e<40$\textdegree). \textbf c, The electron emission angle is restricted to backward scattering ($140<\theta_e<180$\textdegree). The black dots are the experimental data. The error bars represent the standard statistical error. The solid lines are the theoretical results of Approach I and the dashed lines are the results of Approach II (see Methods). The experimental data in \textbf a and \textbf b are normalized such that the maximum intensity is 1; the theory is normalized such that the integrals of the experimental data and the theoretical curves are equal. The normalization factors in \textbf c are identical to those in \textbf b, because here we depict the forward/backward direction of the same distribution.}
	\label{fig3}
\end{figure*}

The observations suggest a two-step model for below-threshold Compton scattering, which is referred to as the $A^2$ approximation (see Methods). The first step is the scattering of the photon at an electron being described by the Thomson cross-section. This step sets the direction and magnitude of the approximate momentum transfer. The second step is the response of the electron wavefunction to this sudden kick, which displaces the bound wavefunction in momentum space. This momentum-shifted electron wavefunction then relaxes to the electronic eigenstates of the ion, where it has some overlap with its initial state and with the bound excited states. However, the fraction that overlaps with the Coulomb continuum  leads to ionization and is observed experimentally. The bimodal electron momentum distribution for small momentum transfer follows naturally from such a scenario. The leading ionizing term in the Taylor expansion of the momentum transfer operator $e^{\boldsymbol Q\cdot\boldsymbol r_e}$ is the dipole operator, with the momentum transfer replacing the direction of polarization. This dipolar contribution, resembling the shape of a $p$ orbital, is the origin of the bimodal electron momentum distribution.

The observed electron momentum distributions are in excellent agreement with the prediction of the $A^2$ approximation shown in Fig.~\ref{fig2}a--c. Note that these theoretical distributions are calculated without any reference to Compton scattering. What is shown is the overlap of the ground state with the continuum (altered by the momentum transfer). Exactly the same distributions are predicted for an attosecond half-cycle pulse (see fig.~2 in ref.~\cite{2}) and identical results are expected for a momentum transfer to the nucleus by neutron scattering \cite{7}.

\begin{figure}
	\centering
	\includegraphics{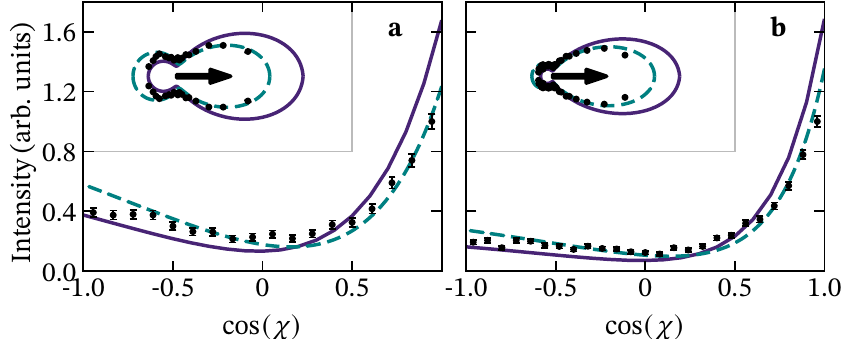}
	\caption{\small\textbf{Fully differential electron angular distributions.}\linebreak \textbf a,\textbf b, The photon scattering angle is $130<\theta<170$\textdegree. Displayed is the cosine of the angle $\chi$ between the outgoing electron and the momentum transfer $\boldsymbol Q$ for electron energies of $1.0<E_e<3.5$\,eV (\textbf a) and $3.5<E_e<8.5$\,eV (\textbf b). Insets show the same data in polar representation, where the arrow indicates the direction of momentum transfer. Black dots are the experimental data, normalized such that the maximum is 1. Error bars represent the standard statistical error. The solid and dashed lines are the theoretical curves resulting from Approach I and Approach II, respectively. The theoretical curves are normalized such that the integral of experiment and theory are equal.}
	\label{fig4}
\end{figure}

Within the $A^2$ approximation, the magnitude of the energy transfer is determined by energy conservation. It is worth mentioning that, under the present conditions, the photon loses only a few percent of its primary energy. Thus the momentum transfer is largely a consequence of the angular deflection of the photon and not a consequence of its change in energy. This can be seen by inspecting the energy distribution of the ejected electron in Fig.~\ref{fig3}a. The electron energy distribution peaks at zero and falls off exponentially. For electron forward emission (Fig.~\ref{fig3}b) it peaks at 11\,eV for photon backscattering, while the backward-emitted electrons for the same conditions are much lower in energy (Fig.~\ref{fig3}c). This also manifests itself in the fully differential cross-section (FDCS) showing the electron angular distribution for fixed electron energy and a fixed photon scattering angle of $150\pm20$\textdegree. These angular distributions (Fig.~\ref{fig4}) show that the intensity in the backward-directed recoil lobe drops strongly with increasing electron energy compared to the intensity in the forward-directed binary lobe. The physics governing the relative strength of the binary and recoil lobes is unveiled by two sets of calculations by comparing theoretical calculations for different initial electron wavefunctions and different final states. First, we use a correlated two-electron wavefunction in the initial state, with outgoing Coulomb waves with charge 1 as the final state. Second, we use a single-active-electron model for the initial state, with a final scattering state in an effective potential (Figs.~\ref{fig3} and \ref{fig4}). We find that the binary peak is similar in all cases. However, the recoil peak is enhanced by more than a factor of two when scattering states in an effective He$^+$ potential are used instead of Coulomb states. This directly supports the mechanistic argument that the recoil peak originates from backscattering of forward-kicked electrons at the parent ion. This backscattering is enhanced due to the increased depth of the effective potential compared to the Coulomb potential close to the origin. The intensity of the recoil peak of both approaches deviates from our experimental data, whereas the shape is predicted correctly by theory. This hints towards the importance of both theoretical approaches (a more detailed discussion is provided in the Methods).

In conclusion, we have shown the first FDCSs for Compton scattering at a gas-phase atom, unveiling the mechanism of near-threshold Compton scattering. Our experimental work shows good agreement with our theoretical models, but further studies with more sophisticated theoretical models are necessary. This work can function as a benchmark measurement for such studies. Coincidence detection of ions and electrons, as demonstrated here, paves the road to exploit Compton scattering for imaging of molecular wavefunctions not only averaged over the molecular axis but also in the body-fixed frame of the molecule. For slightly higher momentum transfers $\boldsymbol Q$, that is, photon energies of $\sim$6\,keV, one can expect the significance of correlations in the scattering states to diminish, simplifying the theoretical description. As has been pointed out recently, measuring the momentum transfer to the nucleus in this case will give access to the Dyson orbitals \cite{11}.\\
\\
\textbf{Acknowledgments}\\
This work was supported by DFG and BMBF. O. Ch. acknowledges support from the Hulubei-Meshcheryakov program JINR-Romania and the RUDN University Program 5-100. Yu. P. is grateful to the Russian Foundation of Basic Research (RFBR) for the financial support under the grant No. 19-02-00014a. S. H. thanks the Direction Generale de la Recherche Scientifique et du Developpement Technologique (DGRSDT-Algeria) for financial support. We are grateful to the staff of PETRA III for excellent support during the beam time. Calculations were performed on Central Information and Computer Complex and heterogeneous computing platform HybriLIT through supercomputer ``Govorun'' of JINR.\\
\\
\textbf{Author contribution}\\
M.K., F.T., S.G., I.V.-P., J.R., S.E., K.B., M.N.P., T.J., M.S.S., and R.D. contributed to the experimental work. S.B., N.E., S.H., O.Ch., Y.V.P., I.P.V., and M.L. contributed to theory and numerical simulations. All authors contributed to the manuscript.\\
\\
\textbf{Data availability}\\
The data that support the plots within this letter are available from the corresponding authors upon reasonable request.\\
\\
\textbf{Code availability}\\
The code that supports the theoretical plots within this letter is available from the corresponding authors upon reasonable request.\\
\\
\textbf{Competing interests}\\
The authors declare no competing interests.
\newpage
\noindent
\textbf{Experimental methods}\\
The experiment was performed at the beamline P04 of the synchrotron PETRA III, DESY in Hamburg with 40-bunch timing mode; that is, the photon bunches were spaced 192\,ns apart. A circularly polarized pink beam was used; that is, the monochromator was set to zero order. To effectively remove low-energy photons from the beam, we put foil filters in the photon beam, namely 980\,nm of aluminum, 144\,nm of copper, and 153\,nm of iron. With this set-up, we suppressed photons $<$100\,eV by at least a factor of $10^{-9}$ and photons $<$15\,eV by at least a factor of $10^{-25}$ (data based on \cite{19}, 9 October 2019, obtained from \href{http://henke.lbl.gov/optical\_constants/filter2.html}{http://henke.lbl.gov/optical\_constants/filter2.html}). The beam was crossed at a {90\textdegree} angle with a supersonic gas jet, expanding through a 30\,{\textmu}m nozzle at 30\,bar driving pressure and room temperature within a COLTRIMS spectrometer. The supersonic gas jet passed two skimmers (0.3\,mm diameter),  so the reaction region had approximate dimensions of 0.2$\times$1.0$\times$0.1\,mm$^3$. The electron side of the spectrometer had 5.8\,cm of acceleration. To increase the resolution, an electrostatic lens and time-of-flight-focusing geometry were used for the ion side to effectively compensate for the finite size of the reaction region. The total length of the ion side was 97.4\,cm. The electric field in the spectrometer was 18.3\,V\,cm$^{-1}$, the magnetic field was 9.1\,G. The charged particles were detected using two position-sensitive microchannel plate detectors with delay-line anodes \cite{20}.\\
\\
\textbf{Theoretical methods}\\
In general, Compton scattering is a relativistic process. In the special case of an initially bound electron, this process may be described by the second-order quantum electrodynamics perturbation terms with exchange in the presence of an external classical electromagnetic field due to the residual ion (see for example \cite{21}). In the low-energy limit of small incoming photon energy $E_1$ compared to the rest energy of an electron, $m_ec^2$, we can apply a non-relativistic quantum-mechanical description~\cite{22,23}. A modern presentation of this approach is provided in ref.~\cite{24}. (In the following, we use atomic units unless stated otherwise; that is, $e=m_e=\hbar=1$.) The energy and momentum conservation laws are of the form
\begin{align}
E_1=E_2 + I_p + E_e + E_{\mathrm{ion}}, \quad \boldsymbol k_1=\boldsymbol k_2+\boldsymbol p_e+\boldsymbol p_{\mathrm{ion}},
\end{align}
where $I_p$ is the ionization potential, $E_e$ ($\boldsymbol{p}_e$) is the energy (momentum) of the escaped electron, $E_{\mathrm{ion}}$ ($\boldsymbol p_{\mathrm{ion}}$) is the energy (momentum) of the residual ion and $E_{1/2}$ ($\boldsymbol k_{1/2}$) are the energies (momenta) of the incoming and outgoing photons, respectively. For the given keV photon energy range, the momenta are of the order $k_i = E_i/c \sim 1\mathrm{\,a.u}$. with the speed of light $c = \alpha^{-1}$ so that the energy of the escaped electron is only a few eV. Given that $M_{\mathrm{ion}} \gg 1$, the ionic kinetic energy $E_{\mathrm{ion}}=\boldsymbol p_{\mathrm{ion}}^2/(2M_{\mathrm{ion}})$ can be neglected. Hence, the photon energy is nearly unchanged and the ratio of photon energy after and before the collision is
\begin{align}
t = \frac{E_2}{E_1} = 1 - \frac{I_p + E_e + E_{\mathrm{ion}}}{E_1} \approx 1.
\end{align}
The transferred momentum from the photon to the atomic system is given by $\boldsymbol Q = \boldsymbol k_1 - \boldsymbol k_2 = \boldsymbol p_e + \boldsymbol p_{\mathrm{ion}}$. The magnitude and direction of the transferred momentum $\boldsymbol Q$ may be expressed as a function of the scattering angle $\theta$ between the incoming and outgoing photon.

Under the above kinematic conditions, the FDCS may be written as
\begin{align}
\frac{\mathrm d\sigma}{\mathrm dE_e \mathrm d\Omega_e \mathrm d\Omega_2} = r^2_e p_e t |M|^2, \label{eq:cs}
\end{align}
with the classical electron radius $r_e$. In this Letter, we use only the so-called $A^2$ (seagull) term from the total second-order Kramers-Heisenberg-Waller matrix element, as is presented, for example, in ref.~\cite{24}:
\begin{align}
M(\boldsymbol Q, \boldsymbol p_e) = (\boldsymbol e_1 \cdot \boldsymbol e_2) \langle \Psi^{(-)}_{\boldsymbol p_e} | \sum^{N}_{j=1}{e^{i\boldsymbol Q \cdot \boldsymbol r_j}} | \Psi_0 \rangle .
\end{align}
Here, $\boldsymbol e_{1/2}$ are the polarization vectors of the incoming and outgoing photons. Initially, the $N$ electrons of the system with positions $\boldsymbol r_j$ are in the bound state $\Psi_0$. Given that in the detection scheme we select singly-ionized helium ions, the final state of the electronic system is a scattering state $\Psi^{(-)}_{\boldsymbol p_e}$ with one electron in the continuum (corresponding to an asymptotic electron momentum $\boldsymbol p_e$) and the other electron remaining bound.

Assuming an unpolarized incoming photon beam and that we do not detect the final polarization state of the outgoing photon, we additionally average over the initial polarization and sum up the probabilities corresponding to both possible orthogonal polarization states. Under these assumptions, the FDCS can be written as
\begin{align}
\frac{\mathrm d \sigma}{\mathrm d E_e \mathrm d \Omega_e \mathrm d\Omega_2} = \left( \frac{\mathrm d\sigma}{\mathrm d\Omega_2} \right)_{\mathrm{Th}} p_e t |M_e|^2 \ .
\end{align}
with the Thomson cross section
\begin{align}
\left( \frac{\mathrm d\sigma}{\mathrm d\Omega_2} \right)_{\mathrm{Th}} = \frac12 r_e^2 (1 + \cos^2 \theta) 
\end{align}
for photons scattered off a single free electron and the electronic matrix element
\begin{align}
M_e(\boldsymbol Q, \boldsymbol p_e) = \langle \Psi^{(-)}_{\boldsymbol p_e} | \sum^{N}_{j=1}{e^{i \boldsymbol Q \cdot \boldsymbol r_j}} | \Psi_0 \rangle . \label{eq:cs2}
\end{align}
From the FDCS the different observables shown in the main text can be calculated.  The $A^2$ approximation resembles the first Born approximation for scattering of a fast particle on an atom, for example $(e,2e)$ ionization by electron impact~\cite{4}. Therefore, the observed effects have an analogous interpretation and can be described in familiar terms. However, the Compton ionization has some advantages compared to traditional methods such as $(e,2e)$ ionization: (1) the contribution of other second-order terms is very small so that the $A^2$ approximation is often accurate; (2) the photon has no charge so that we only need to consider the evolution of the field-free system of charged particles; (3) the transferred momentum $\boldsymbol Q$ can vary in a wide range so that different regimes are accessible.

Compton scattering by a bound electron is a sequential process and may be divided into two steps. In the first, the incoming photon is captured by a bound electron. Afterwards, this dressed system evolves in time so that a photon is emitted and an electron escapes. In the $A^2$ approximation, the second photon is emitted immediately after the absorption so that this short photon scattering process can be effectively interpreted as a `kick' of the electronic bound-state distribution by the transferred momentum~$\boldsymbol{Q}$. The corresponding scattering probability is described by the Thomson formula. The `kicked', field-free atomic system evolves in time. One part of the boosted wave function remains bound, while the other part is set free in the continuum and causes ionization. In principle, the time evolution including the interaction between electrons and their possible correlation is implicitly contained in the scattering state $\Psi^{(-)}_{\boldsymbol p_e}$ in equation (\ref{eq:cs2}). However, the calculation of fully-correlated scattering states is beyond the scope of this work.

To calculate the electronic matrix elements, com\-ple\-men\-ta\-ry approaches have been used: The first model (Approach I) describes both electrons and takes into account correlation in the ground state, but uses Coulomb waves as scattering states. In contrast, the second model (Approach II) uses a single-active-electron description, but includes accurate one-electron scattering states.\\
\\
\textbf{Approach I: model with correlated ground state}\\
In the first approach, both electrons of the helium atom are explicitly treated such that the `direct' ionization of the `kicked' electron as well as the `shake-off' (that is, ejection of the unkicked electron) are considered. In equation (\ref{eq:cs2}), the initial state is given by a correlated symmetric two-electron ground state $\Psi_0(\boldsymbol r_1, \boldsymbol r_2)$, obtained from \cite{23}. To approximate the final state, the main idea is that one electron remains bound in the ionic ground state given by
\begin{align}
\psi_0^{\text{He}^+}(\boldsymbol r) = \sqrt{\frac{8}{\pi}} \, e^{-2r}
\end{align}
and the free electron may be approximated by Coulomb wavefunctions
\begin{align}
\psi_{\boldsymbol p_e}^{C}(\boldsymbol r) = \sqrt{\frac{e^{-\pi \zeta}}{(2\pi)^{3}}} \, \Gamma(1-i\zeta) e^{i\boldsymbol p_e \cdot \boldsymbol r} \, _1F_1 (i\zeta, 1, -ip_e r - i\boldsymbol p_e \cdot \boldsymbol r)
\end{align}
with $\zeta = -1/p_e$ and $_1F_1$ being the confluent hypergeometric function. Because the correct scattering states $\Psi^{(-)}_{\boldsymbol p_e}(\boldsymbol r_1, \boldsymbol r_2)$ have to be orthogonal to the initial bound states, the resulting symmetrized final state
\begin{align}
\tilde{\Psi}_{\boldsymbol p_e}^{(-)}(\boldsymbol r_1, \boldsymbol r_2) = \frac{1}{\sqrt{2}} \left[ \psi_{\boldsymbol p_e}^{C} (\boldsymbol r_1) \psi_0^{\text{He}^+}(\boldsymbol r_2) + \psi_{\boldsymbol p_e}^{C} (\boldsymbol r_2) \psi_0^{\text{He}^+}(\boldsymbol r_1) \right]
\end{align}
is afterwards explicitly orthogonalized with respect to the initial state $\Psi_0$ such that the electronic matrix elements of equation ($\ref{eq:cs2}$) read
\begin{align}
M_e(\boldsymbol Q, \boldsymbol p_e) =& \langle \Psi_{\boldsymbol p_e}^{(-)} | e^{i\boldsymbol Q \cdot \boldsymbol r_1} + e^{i\boldsymbol Q \cdot \boldsymbol r_2} | \Psi_0 \rangle \nonumber \\
=& \langle \tilde{\Psi}_{\boldsymbol p_e}^{(-)} | e^{i\boldsymbol Q \cdot \boldsymbol r_1} + e^{i\boldsymbol Q \cdot \boldsymbol r_2} | \Psi_0 \rangle -\nonumber \\ & \langle \tilde\Psi_{\boldsymbol p_e}^{(-)} | \Psi_0 \rangle \langle \Psi_0 | e^{i\boldsymbol Q \cdot \boldsymbol r_1} + e^{i\boldsymbol Q \cdot \boldsymbol r_2} | \Psi_0 \rangle
\end{align}
\textbf{Approach II: single-active-electron model}\\
In the second approach only the `kicked' electron may escape, while the other electron stays frozen at the core. To model the influence of the remaining electron on the escaping electron, we use a single-active-electron effective potential \cite{24}. This potential has an asymptotic charge of $Z=2$ for $r\rightarrow0$, which is screened by the second electron such that asymptotically for large $r$, it reaches $Z=1$. The one-electron ground state $\psi_0$ and the one-electron continuum state $\psi_{\boldsymbol p_e}^{(-)}$ with incoming boundary conditions are calculated numerically via solving the radial Schr\"odinger equation. Hence, the electronic matrix element in equation (\ref{eq:cs2}) is approximated as
\begin{align}
M_e(\boldsymbol Q, \boldsymbol p_e) = \sqrt{2} \, \langle \psi_{\boldsymbol p_e}^{(-)} | e^{i\boldsymbol Q \cdot \boldsymbol r} | \psi_0 \rangle.
\end{align}
This expression is calculated using a plane wave expansion of $e^{i\boldsymbol Q\cdot \boldsymbol r}$ and an expansion of the scattering states $\psi^{(-)}_{\boldsymbol p_e}$ in terms of spherical harmonics.

Both approaches use two main approximations. (1) The final scattering states are not the exact fully correlated states. This leads to deviations in the low energy region at the recoil peak. In particular, `shake-off' and `shake-up' processes are not fully included. To some extent, correlations are included due to the orthogonalization in Approach I and the effective potential that was used in Approach II. However, we believe that including correlations in the final state in a more systematic way is more important than in the ground state. (2) The state of the residual ion has not been resolved in the experiment. In Approach I, we assumed that the bound electron remains in the ground state of the ion whereas it is simply frozen in the ground state of the atom in Approach II. We expect that this works well for the binary peak (forward direction), but not for the recoil peak (backward direction). To improve the calculations, ionization in different channels corresponding to excited states of the residual ion need to be considered.
\bibliographystyle{apsrev}

\end{document}